# Biomimetic wet-stable fibres via wet spinning and diacid-based crosslinking of collagen triple helices

M. Tarik Arafat,[a,b] Giuseppe Tronci,[a,b,*] Jie Yin,[a,b] David J. Wood,[b] Stephen J. Russell[a]

[a] Nonwovens Research Group, Centre for Technical Textiles, School of Design, University of Leeds, UK.

[b] Biomaterials and Tissue Engineering Research Group, School of Dentistry, St James University Hospital, University of Leeds, UK.

*corresponding author
Giuseppe Tronci (e-mail: G.Tronci@leeds.ac.uk)

**Graphical abstract:**

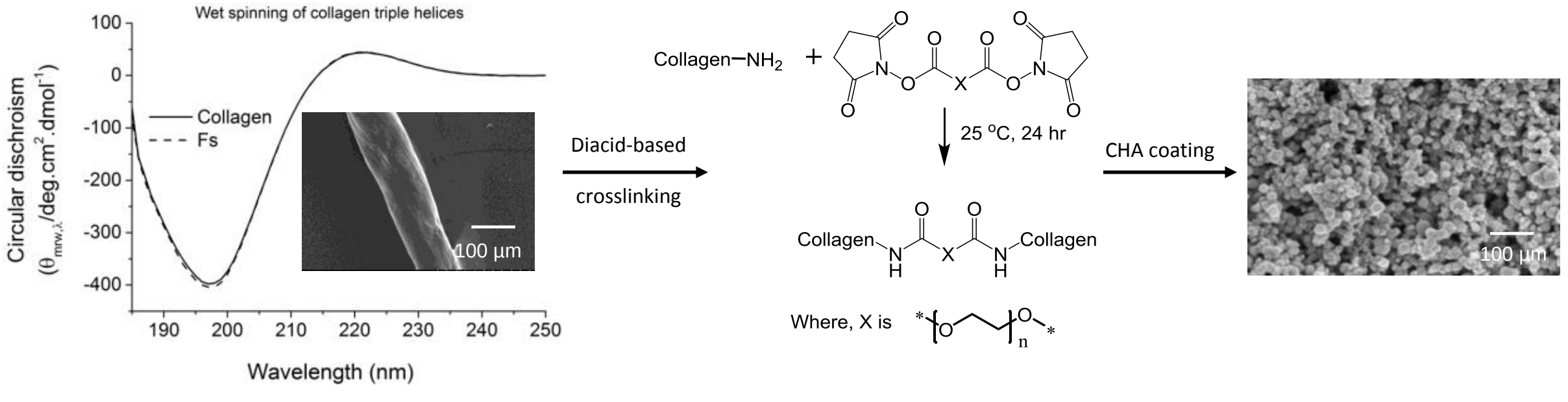

**Abstract:**

One of the limitations of electrospun collagen as bone-like fibrous structure is the potential collagen triple helix denaturation in the fibre state and the corresponding inadequate wet stability even after crosslinking. Here, we have demonstrated the feasibility of accomplishing wet-stable fibres by wet spinning and diacid-based crosslinking of collagen triple helices, whereby fibre ability to act as bone-mimicking mineralisation system has also been explored. Circular dichroism (CD) demonstrated nearly complete triple helix retention in resulting wet-spun fibres, and the corresponding chemically crosslinked fibres successfully preserved their fibrous morphology following 1-week incubation in phosphate buffer solution (PBS). The presented novel diacid-based crosslinking route imparted superior tensile modulus and strength to the resulting fibres indicating that covalent functionalization of distant collagen molecules is unlikely to be accomplished by current state-of-the-art carbodiimide-based crosslinking. To mimic the constituents of natural bone extra cellular matrix (ECM), the crosslinked fibres were coated with carbonated hydroxyapatite (CHA) through biomimetic precipitation, resulting in an attractive biomaterial for guided bone regeneration (GBR), e.g. in bony defects of the maxillofacial region.

## 1. Introduction

Bone tissue engineering (TE) is a fascinating field of research with substantial focus on delivering materials that mimic natural constituents of bone. Especially in the maxillofacial context, natural polymers, such as collagen, have been widely employed for the design of biomimetic membranes for guided tissue regeneration (GBR) [1], aiming to accomplish selective, endogenous bone tissue growth into a defined space maintained by tissue barriers [2]. The fabrication of tissue-mimicking biomaterials is a key to successful GBR. In this regards, the extracellular matrix (ECM), which rules the structure, properties and functions of bone and comprises both non-mineralized organic and mineralized inorganic components should be greatly considered [3]. The main organic component of bone is type I collagen, which forms more than 90% of its organic mass [4, 5]. Although the architectures and roles of these collagens vary widely, they all comprise triple helix bundles at the molecular scale, where the collagen molecule consists of a right-handed triple helix composed of about 1000 amino acids with two identical $\alpha_1$(I) and $\alpha_1$(II) chains, and one $\alpha_2$(I) chain. These three chains, staggered by one residue relative to each other, are supercoiled around a central axis in a right-handed manner to form the triple helix which is around 300 nm in length and 1.5 nm in diameter [4, 6-9]. The strands are held together mostly by hydrogen bonds between adjacent -CO and -NH groups [10]. On the other hand what defines bone as a mineralized tissue is the deposition of inorganic carbonated apatite, and this mineral deposition is mainly accomplished by the precipitation of the apatite phase. Initially the precipitation occurs via matrix vesicle nucleation alone but, ultimately, requires collagen structures [5]. Thus, carbonated apatite coated collagen fibres, which are obtained from the assembly of collagen triple helices and are mineralized with apatite, are considered as the building blocks of ECM. In order to mimic this unique material constituents present *in vivo*, the formation of triple-helical collagen fibres and respective biomimetic mineralization with carbonated apatite has been addressed in this paper.

Owing to the excellent biological features and physiochemical properties of collagen, it has been among the most widely used biomaterials for biomedical applications [6, 10-12] particularly when delivered in the form of gels, films, injectables and coatings [13-17]. To further extend the utility of collagen for use in medical devices and to address issues such as fixation in defect sites and stability for TE and GBR, the ability to manufacture mechanically robust fibres and fabrics is important. However, avoiding denaturation of the native triple helical structure and the instability of regenerated collagen materials in the hydrated state still remains highly challenging [18]. Electrospinning, which has been the main collagen fibre manufacturing process for use in TE, has limitations. The organic reagents required to prepare collagen electrospinning solutions such as those based on fluoroalcohols [19, 20] are known to be highly toxic and partially denature the native structure of collagen [18, 20]. To address this issue non-toxic solvents such as PBS/ethanol or acetic acid have been successfully introduced [21, 22]. Among the different fibre manufacturing processes available, wet spinning has the potential to convert biomolecules into fibres without need for high voltage during manufacture and is less likely to be associated with denaturation [23, 24]. Wet spinning was developed by the textile industry in the early 1900s as a means of producing man-made fibres such as viscose rayon. This fibre spinning technology is based on non-solvent induced phase separation, whereby polymer dope solutions are extruded through a spinneret into a non-solvent coagulation bath in which liquid polymer streams turn into solid filaments. In this context, using wet spinning to manufacture fibres from collagen triple helices suspension utilising non-toxic solvents such as acetic acid, whilst avoiding the addition of any synthetic phase, could be promising.

Many studies have been conducted to stabilize collagen fibres via covalent crosslinking [25-27]. The most popular crosslinking method is via carbodiimide, especially 1-ethyl-3-(3-dimethylaminopropyl) carbodiimide hydrochloride (EDC), often used in the presence of N-hydroxysuccinimide (NHS). This method leads to activation of carboxylic functions and subsequent formation of amide net-points between amino and carboxylic functions of collagen [25-27]. One of the main reason of the popularity of EDC/NHS treatment is its compatibility compared to the commonly used bifunctional crosslinker, glutaraldehyde (GTA) [27]. However, EDC/NHS mostly links carboxylic acid and amino groups that are located within 1.0 nm of each other, which eventually means functional groups that are located on adjacent collagen molecules are too far apart to be bridged by carbodiimide [28, 29]. A high molecular weight diacid could be a potential candidate to crosslink collagen fibres in presence of EDC/NHS. To address this hypothesis, we have evaluated the utility of 1, 3-phenylenediacetic acid (Ph) and poly(ethylene glycol) bis(carboxymethyl) ether (PEG) as bifunctional crosslinkers of varied segment length. Ph has been used to form biocompatible stable collagen hydrogels [30]; and PEG, which is an FDA approved chemical for several medical and food industries [31], has also been used to stabilize collagen [32-34] and polysaccharides [35].

Besides the nonmineralized collagen component, another important constituent of bone ECM is a mineralized inorganic component. Therefore, in order to develop attractive biomaterials for GBR, a composite of collagen and mineralized component should be considered. Initially the apatite was assumed to be hydroxyapatite, however, due to the presence of significant amount of carbonate, it is better defined as carbonate hydroxyapatite (CHA) [36]. Composites of biopolymer and hydroxyapatite (HA) were investigated in this regard, however, in a composite the functionality of HA is reduced due to the masking of apatite particles by biopolymers [37, 38]. Therefore, a biomimetic coating of apatite on a collagen template can be considered an efficient approach [39, 40]. By mimicking the natural biomineralization process Kokubo *et al.* first reported the use of simulated body fluid (SBF) for biomimetic growth of apatite coatings on bioactive CaO–SiO2 glasses [41]. SBF has also been used to form apatite coating on collagen formulations [39, 42, 43]. However, SBF possesses limitations, particularly the time consuming nature of the process, necessity of a constant pH and constant replenishment to maintain super-saturation for apatite crystal growth [37, 39, 44, 45]. Therefore, an alternative, simple and efficient approach for biomimetic coating on collagen fibre is needed.

The aim of the current study was to achieve wet-stable fibres via wet spinning and covalent crosslinking of collagen triple helices, and also to mimic the constituents of natural bone ECM through the precipitation of CHA on the as-formed wet-stable fibres. Ph and PEG were used as diacids of varied molecular weight to increase the likelihood of crosslinking distant collagen molecules was compared to the state-of-the-art zero length crosslinker EDC. Finally, CHA was coated through a biomimetic precipitation process on the crosslinked wet-spun fibres.

## 2. Materials and methods

### 2.1 *Materials*

1-ethyl-3-(3-dimethylaminopropyl) carbodiimide hydrochloride (EDC) and N-hydroxysuccinimide (NHS) and 1, 3 phenylenediacetic acid (Ph) were purchased from Alfa Aesar. 2,4,6-trinitrobenzenesulfonic acid (TNBS), acetic acid ($CH_3COOH$), calcium chloride ($CaCl_2$), phosphoric acid ($H_3PO_4$), sodium carbonate ($Na_2CO_3$), Potassium phosphate dibasic trihydrate ($K_2HPO_4.3H_2O$), poly(ethylene glycol) bis(carboxymethyl) ether (PEG) and Dulbecco's Phosphate Buffered Solution (PBS) were purchased from Sigma Aldrich. Tissue culture media, Dulbecco's modified Eagle's medium (DMEM), fetal calf serum (FBS) and penicillin-streptomycin (PS) were purchased from Gibco. CellTiter® 96 AQueous one solution cell proliferation assay were purchased from Promega UK Ltd.

### *2.2 Isolation of type I collagen from rat tail tendons*

Type I collagen was isolated through acidic treatment of rat tail tendons as described in previous papers [30, 46]. In brief, frozen rat tails were thawed in 70% ethanol for about 15 min. Individual tendons were pulled out of the tendon sheath and placed in 50 ml of 17.4 mM acetic acid solution for each rat tail at 4 ℃ in order to extract collagen. After three days the supernatant was centrifuged at 10000 r·min$^{-1}$ for half an hour. The mixture was then freeze-dried in order to obtain type I collagen. The resulting product showed only the main electrophoretic bands of type I collagen during sodium dodecyl sulphate-polyacrylamide gel electrophoresis (SDS-page) analysis [46].

Collagen dissolved in 17.4 mM acetic acid was studied by reading the optical density at 310 nm using a photo spectrometer. 0.25% (w/v) of collagen was mixed in 17.4 mM acetic acid to carry out the experiment. The optical density of the hydrolysed fish collagen solution was also studied as a control, given the absence of triple helix in respective CD spectrum in the same conditions [47].

2.3 *Formation of wet-spun fibres (Fs)*

To accomplish Fs from collagen triple helices, collagen was dissolved in 17.4 mM acetic acid in different concentrations (0.8 – 1.6% wt/vol) at 4 ℃ overnight. Resulting collagen suspensions were transferred into a 10 ml syringe having 14.5 mm internal diameter. The collagen suspensions were then ejected from the syringe through a syringe pump at a dispensing rate of 12 ml·hr$^{-1}$ with the syringe tip submerged in a coagulation bath containing 1 litre of pure ethanol at room temperature. The as-formed fibres were then removed from the ethanol and dried separately at room temperature.

2.4 *Crosslinking of Fs*

Fs (3 mg) were crosslinked either via Ph, PEG or EDC. To activate Ph with NHS, 5 or 10 mg of Ph (corresponding to either 25 or 50 fold molar excess of Ph with respect to collagen lysines) were dissolved in 3 ml ethanol solution containing either 0.1 or 0.2 M EDC and an equimolar content of NHS, respectively. PEG-crosslinked fibres were obtained following the previous protocol with the only difference that either 15 or 31 mg PEG was used instead of Ph. Ultimately, crosslinked control groups were obtained by incubating fibres with just EDC and NHS (both with either 0.1 or 0.2 M concentration in ethanol).

Crosslinked fibres are denoted as 'F-XXXYY', where 'XXX' indicates the type of crosslinking treatment, i.e. via either EDC/NHS (EN), Ph or PEG. 'YY' identifies the molar content of the EDC/NHS, Ph or PEG in the reacting mixture. Therefore 'YY' is either '1' (in the case of the reacting mixture with the highest concentration of either EN, Ph or PEG) or '0.5' (in the case of reacting mixture at half the concentration).

2.5 *2,4,6-Trinitrobenzenesulfonic acid (TNBS) assay*

2,4,6-trinitrobenzenesulfonic acid (TNBS) colorimetric assay was used to determine the degree of crosslinking in reacted fibres [48]. Briefly, 4 wt% $NaHCO_3$ (pH 8.5) and 1 ml of 0.5 wt% TNBS solution were added to 11 mg of dried crosslinked fibres; and the temperature of the mixture raised to 40 ℃ for 4 hr under mild shaking. 3 ml of 6 M HCL solution were added to the mixture, and the mixture was kept at 80 ℃ for 1 hr. Blank samples were prepared for all the groups following the above procedure, except that the HCl solution was added before the addition of TNBS. The content of free amino groups and degree of crosslinking (C) were calculated as follows:

$$\frac{moles\ (Lys)}{g\ (collagen)} = \frac{2\ \times 0.02\ \times Absorbance\ (346\ nm)}{1.46\ \times 10^4\ \times x\ \times b}$$

$$Degree\ of\ crosslinking, C = \left(1 - \frac{moles(Lys)_{crosslinked}}{moles(Lys)_{collagen}}\right) \times 100$$

Where, *Absorbance (346 nm)* is the absorbance value at 346 nm, *$1.46\times10^4$* is the molar absorption coefficient of 2,4,6-trinitrophenyl lysine ($l\cdot mol^{-1}\cdot cm^{-1}$), *0.02* is the solution volume (l), *2* is the dilution factor, *x* is the weight of sample (g), *b* is the cell path length (1 cm); $moles(Lys)_{crosslinked}$ and $moles(Lys)_{collagen}$ represent the lysine molar content in crosslinked and native collagen, respectively. For each sample composition two replicas were used.

2.6 *Biomimetic coating on fibres*

PEG-crosslinked fibres (F-PEG0.5) were coated as described previously [37, 38]. Initially $CaHPO_4$ was coated on the F-PEG0.5, to act as nucleation sites, by dipping the sample into calcium chloride solution and dipotassium hydrogen phosphate solution alternately [49, 50]. In brief, F-PEG0.5 was dipped in $CaCl_2$ aqueous solutions (20 mL, 0.2 M) for 10 min, and then dipped in de-ionized water for 10 s followed by air drying for 2 min. The treated samples were subsequently dipped in $K_2HPO_4$ aqueous solutions (20 mL, 0.2 M) for 10 min, and then dipped in de-ionized water for 10 s followed by air drying for 2 min. The whole process was repeated three times. The $CaHPO_4$ coated F-PEG0.5 were then kept for subsequent processing. Meanwhile, $CaCl_2$ (10 ml, 0.1 M) and $H_3PO_4$ (6 ml, 0.1 M) with a Ca/P ratio of 1.66 were simultaneously added drop wise into $CH_3COOH$ (20 ml, 0.1 M) while stirring. $Na_2CO_3$ (18 ml, 0.1 M) with the molar ratio of $CO_3^{2-}/PO_4^{3-}$ was gradually added into the solution after 30 min of stirring. The mixture was further stirred for 30 min before titrating it to pH 9 using NaOH (0.1 M). At this stage $CaHPO_4$ coated samples were immersed in the solution; this was to make sure collagen was not exposed to acidic environment, as prolonged exposure can denature the material. Under the coating conditions carboxyl and amine groups present in the F-PEG0.5 may become charged $COO^-$ and $NH_3^+$, respectively which can also promote nucleation. CHA coated F-PEG0.5 (F-PEG0.5-CHA) were collected after the solution was aged for 3 hr. Finally, samples were washed with de-ionized water and freeze dried.

2.7 *Characterisation*

*2.7.1 Mechanical properties*

Tensile modulus and tensile strength of the samples were measured using a Zwick Roell Z010 testing system with 10 N load cell at a rate of 0.03 mm·$s^{-1}$. The testing was carried out in a controlled environment with a room temperature of 18 ℃ and relative humidity of 38%. The gauge length was 5 mm. Ten individual fibre samples were tested from each group and measurements were reported as mean ± standard deviation. To measure the tensile modulus and strength in hydrated conditions, samples were immersed in phosphate buffer solution (PBS) at 37 ºC for 24 hr prior to tensile testing.

*2.7.2 Viscosity*

The viscosities of collagen suspensions having different concentrations (0.8 – 1.6% wt/vol) were studied using a bench top Brookfield DV-E Viscometer (Brookfield Engineering Laboratories, Inc., Middleboro, MA, USA) with different spindles (s34 and s25). The experiment was conducted at room temperature with speeds ranges from 1 to 100 rpm. The amount of collagen suspension used was 9.4 ml and 16.1 ml for spindle s34 and s25, respectively, in accordance with the manufacturer's protocol.

*2.7.3 Surface morphology and chemical structure*

Fibre surface morphology was observed using a Hitachi SU8230 FESEM. Samples were coated with at a beam intensity of 10 kV after gold sputtering using a JFC-1200 fine sputter coater. The chemical structure of the coated fibres was analysed using an Oxford instrument X-man attached to the FESEM.

*2.7.4 Circular dichroism (CD)*

Solutions of native collagen and wet spun fibres were prepared by dissolving 1 mg dry material in 5 ml HCL (10 mM) and stirring overnight at room temperature. The prepared solutions were used to acquire CD spectra using a Jasco J-715 spectropolarimeter. Sample solutions were collected in quartz cells of 1.0 mm path length and CD spectra were obtained with 20 nm·min$^{-1}$ scanning speed and 2 nm band width. A spectrum of the 10 mM HCl control solution was subtracted from each sample spectrum.

*2.7.5 Swellability*

Swelling tests of the fibres were performed by incubating dry samples in 50 ml deionized water for 24 hr at 37 ℃. The water equilibrated samples were paper blotted and weighed to get the swollen sample weight. The weight based swelling ratio ($S_R$) was calculated as follows:

$$S_R = \frac{wt_s - wt_d}{wt_d} \times 100$$

Where, $wt_s$ and $wt_d$ are swollen and dry sample weight, respectively. Five replicas were used for each sample group, and results were reported as mean ± standard deviation.

*2.7.6 Hydrolytic Degradation*

The hydrolytic degradation of the crosslinked samples were tested by incubating dry samples in 50 ml of PBS for 7 days at 37 ℃. Retrieved samples were rinsed with distilled water, freeze dried and weighed. The mass change ratio ($M_R$) due to hydrolytic degradation was calculated as follows:

$$M_R = \frac{wt_h - wt_d}{wt_d} \times 100$$

Where, $wt_h$ and $wt_d$ are dry sample weight of after and before PBS incubation, respectively. Five replicas were used for each sample group, and results were reported as mean ± standard deviation.

*2.7.7 Thermal Stability*

Differential scanning calorimetry (DSC) on fibres was carried out using a thermal analysis 2000 system and 910 differential scanning calorimeter cell base (TA instruments) in the range of 10 – 120 ℃ with a heating rate of 10 ℃·$min^{-1}$. The DSC cell was calibrated using indium with a heating rate of 10 ℃·$min^{-1}$ under 50 $cm^3 \cdot min^{-1}$ nitrogen atmosphere. Sample weight for each measurement was 3-5 mg.

*2.7.8 Fourier Transform Infrared Spectroscopy*

Attenuated total reflectance Fourier transform infrared spectroscopy (ATR-FT-IR) analysis of native, crosslinked and coated fibres was performed in a Perkin-Elmer Spectrum BX over a range of 600–1800 $cm^{-1}$ at resolution of 2 $cm^{-1}$ to study the chemical structure of the coatings.

*2.7.9 Extract Cytotoxicity*

An extract cytotoxicity assay was conducted in order to further investigate the material compatibility with L929 cells via both quantitative MTS assay and qualitative cell morphology observations (EN DIN ISO standard 10993–5). For each group 1 mg of fibres were sterilized with 70% ethanol for 30 min and then further sterilized under UV light for 15 minutes followed by 30-min incubation in PBS. Resulting samples were weighed in sterile conditions and incubated in completed (10% FCS, 1% PS) DMEM (10 µL completed DMEM· $\mu g^{-1}$ hydrated fibre) at 37 ℃ for 72 hrs. At the same time, L929 mouse fibroblasts ($10^5$ cells · $mL^{-1}$) were seeded on to a 96-well plate (100 µL cell suspension per well) for 24 hours to enable cell confluence. After that, cell culture medium was replaced with sample extract into each well and cells cultured for 48 hrs. Each group of samples had 8 replicas. Non crosslinked fibres were used as positive control, while dimethyl sulfoxide (DMSO) was used as negative control. Transmitted light microscopy was used to monitor cell morphology, while CellTiter® 96 AQueous, which is a colorimetric cytotoxicity assay, was used to quantify the number of viable cells in proliferation as per manufacturer's protocol. The absorbance of the samples was measured at 490 nm.

*2.8 Statistical analysis*

All the data presented are expressed as mean ± standard deviation. An unpaired student's t-test was used to test the significance level of the data. Differences were considered statistically significant at $p<0.05$.

## 3. Results and Discussion

3.1 *Fabrication of fibres*

Fig. 1 provides a schematic representation of the synthetic approach pursued in this study in order to accomplish biomimetic wet-stable fibres. Micron-scale fibres were obtained via wet-spinning of collagen triple helices, which were subsequently reacted with different di-acid crosslinkers in order to preserve the fibre morphology in physiological conditions. Ultimately resulting material was mineralized with CHA through a biomimetic precipitation process, aiming to mimic the constituents of bone tissue.

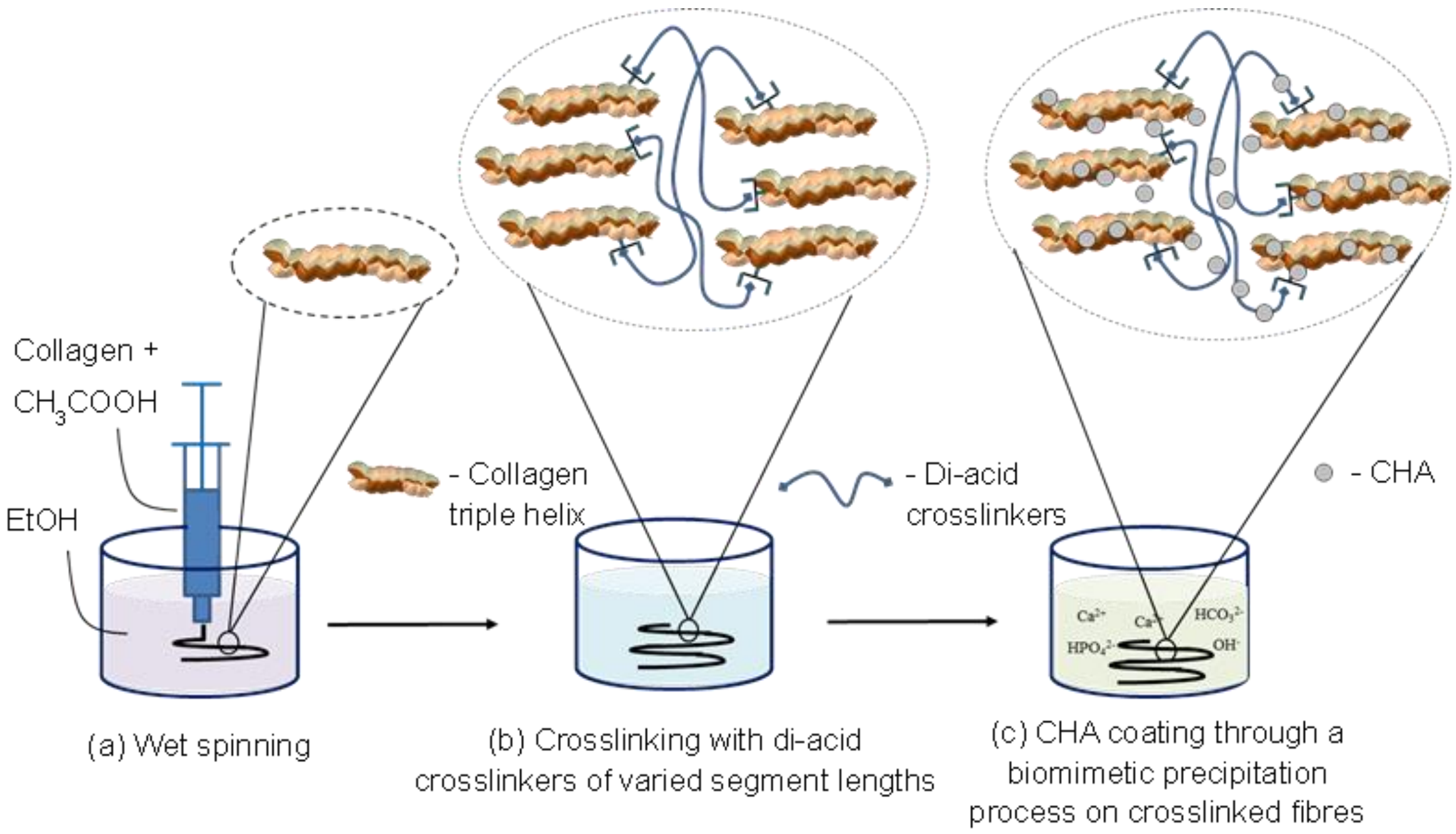

**Fig. 1.** Formation of biomimetic wet-stable fibres. (A) A suspension of collagen triple helices is wet-spun against ethanol, leading to the formation of micron-scale fibres. (B) Wet-spun fibres are reacted with varied diacids, resulting in a molecular network of covalently-crosslinked collagen triple helices. (C) Wet-stable covalently-crosslinked fibres are mineralised with CHA through a biomimetic process in order to obtain bone-mimicking fibrous systems.

From a biological perspective, wet spinning is an attractive method for the formation of biomimetic fibres, because, unlike electrospinning, there is no need to subject polymer solutions to high voltage, resulting in negligible risk of denaturation of collagen macromolecules [23]. At the same time, wet-spinning may also allow for a systematic control in fibre morphology and orientation depending on selected experimental conditions, which is only partially accomplished via electrospinning. In wet spinning, polymer concentration is known to affect fibre morphology and resulting material properties [24]. Here, viscosity measurements on different collagen suspensions as well as dry tensile tests and SEM were carried out on collected fibres in order to identify the range of collagen concentrations enabling formation of homogeneous fibres. Fibres could not be formed with collagen suspensions as low as 1% (w/v), whilst the concentrations range of 1.2-1.6% (w/v) collagen was found to be suitable for fibre formation. Representative tensile data for resulting single fibres are shown in Fig 2. Fibres fabricated from 1.2% (w/v) collagen suspension displayed improved tensile mechanical properties with respect to fibres formed from either 1.4 or 1.6% (w/v) collagen suspensions, whereby a tensile modulus and tensile strength of 2200 MPa and 71.76 MPa, respectively, was observed. Similarly to the decrease in tensile modulus and strength, the tensile strain of fibres obtained from suspensions with increased collagen concentration was also observed to decrease [Fig 2(b)].

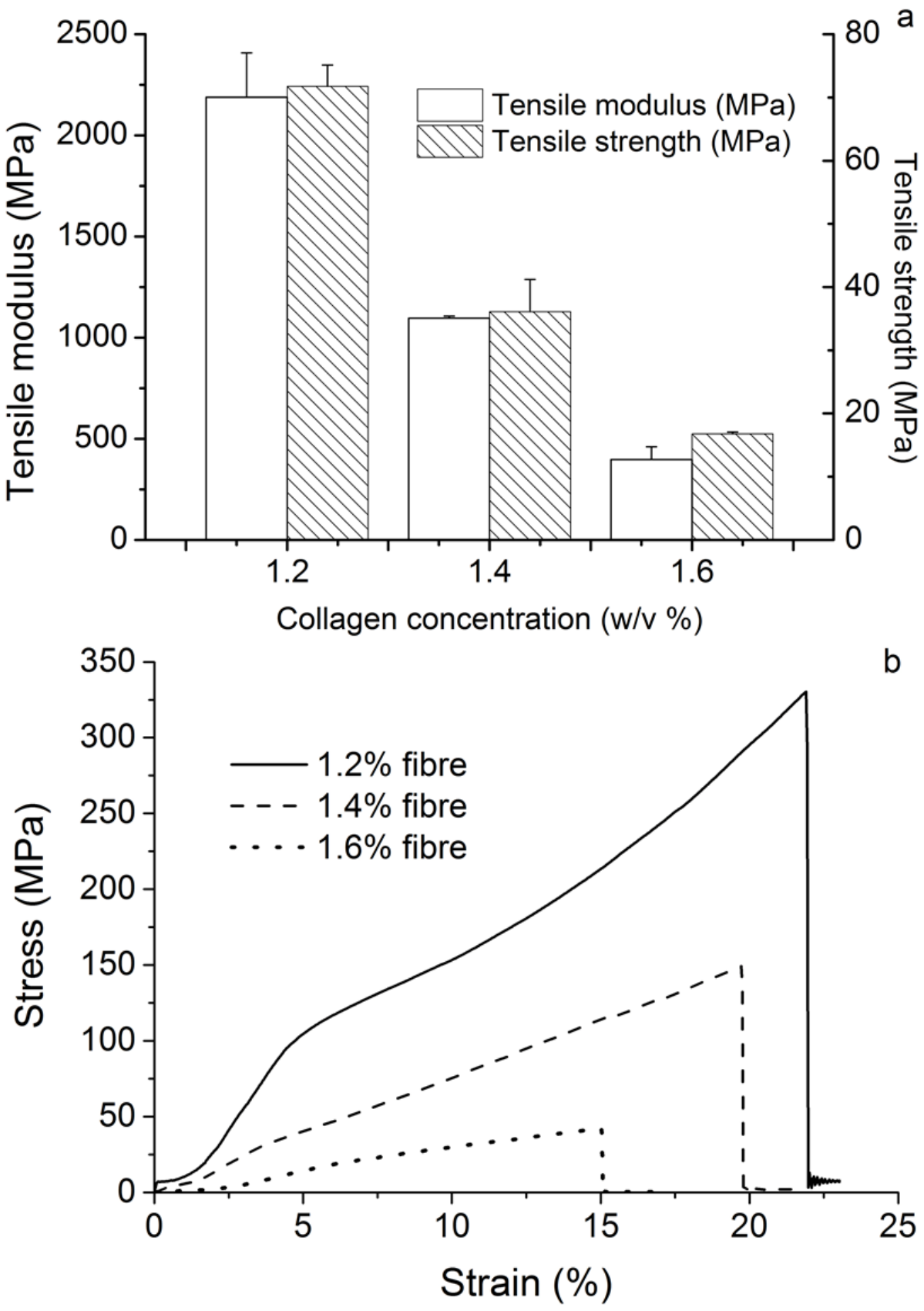

**Fig. 2.** Variation of tensile modulus and tensile strength (a) and corresponding exemplary stress-strain curves (b) of fibres obtained via wet-spinning of varied collagen suspensions. Fibres wet-spun from 1.2% (w/v) collagen suspension exhibited statistically enhanced tensile strength and modulus in comparison to fibres obtained from suspensions with increased collagen concentration. Tensile modulus and strength of all the groups are significantly different to each other ($p < 0.05$, t-test).

This inverse relationship between fibre tensile properties and wet spinning collagen suspension concentration can be rationalised by the assumption that at higher concentration collagen molecules have less chance to align during suspension spinning, leading to reduced tensile modulus, strength and strain [51]. This assumption is supported by the viscosity data (Fig. 3) for various wet-spinning collagen suspensions, which indicated a sharp increase in the shear stress at low shear rates when the collagen suspension concentration was increased from 0.8 to 1.6% (w/v). This increase in suspension viscosity is expected to inhibit the movement and axial orientation of collagen molecules (as a result of shear during extrusion), thereby inevitably resulting in randomly aligned collagen molecules at the molecular scale and defected fibres at the micro-scale.

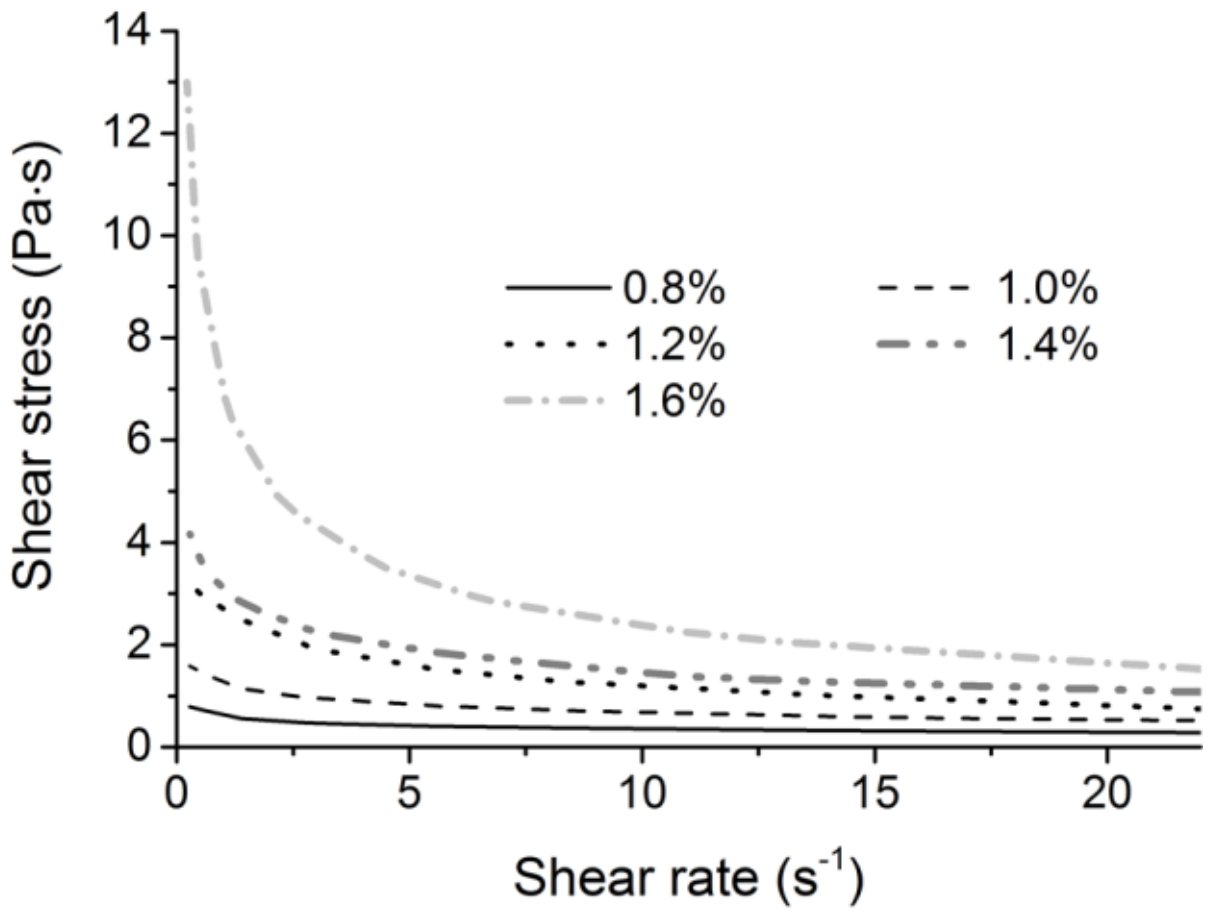

**Fig. 3.** Shear stress vs. shear rate curves obtained via viscosity measurements on wet spinning collagen suspensions with varied collagen concentrations (w/v).

The SEM image of fibres (Fig. 4) revealed that fibres obtained at collagen concentrations in the range of 1.4-1.6% (w/v) were longitudinally striated, a characteristic that became more evident at the highest concentration (Fig. S1). The mean diameter of the fibres also increased with increasing collagen concentration.

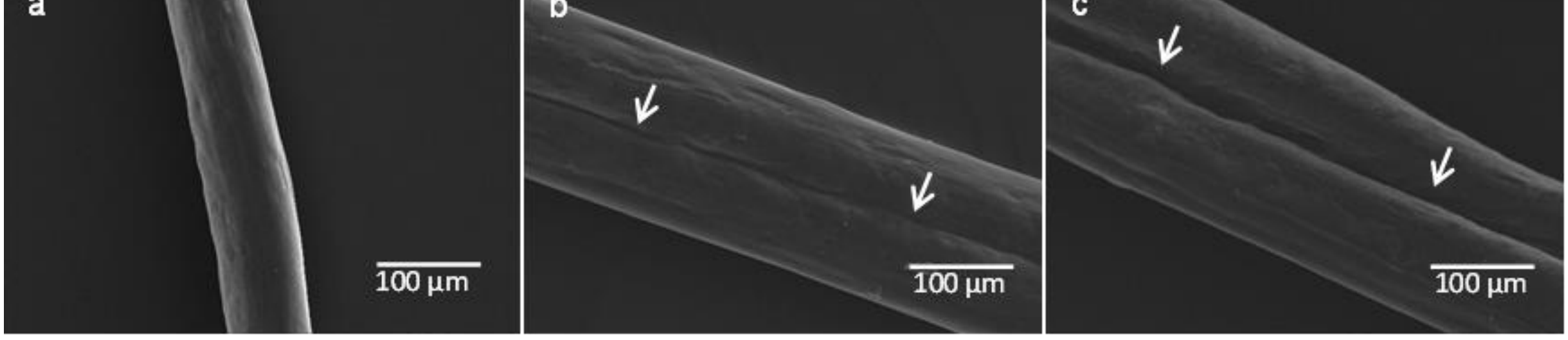

**Fig. 4.** SEM images of fibres wet spun from suspensions with varied collagen concentration: (a) 1.2%, (b) 1.4% and (c) 1.6% collagen (w/v) suspensions. Surface striations (marked with arrow) are observed in fibres deriving from wet-spinning suspensions with increased collagen concentrations.

Clearly, an in-house wet spinning apparatus was employed in this study, whereby drawing (stretching) of the as-spun fibres was not carried out. By appropriate adjustment of spinning conditions, drawing and fibre dimensions, mechanical properties could be substantially modified.

Besides the formation of uniform fibres, a major challenge in manufacturing collagen fibres is the potential denaturation of collagen triple helices following fibre formation. Collagen denaturation can be caused by the manufacturing conditions and/or the solvent associated with the manufacturing process [18]. Optical density was investigated following dissolution of isolated collagen and hydrolysed fish peptide (as control) in 17.4 mM acetic acid, to verify the presence (absence) of collagen triple helix on collagen (hydrolysed fish collagen) solution (in accordance with CD data). The optical density values (Fig. S2) at 310 nm of collagen and hydrolysed fish peptides were found to be 0.9 and 0.02, respectively, which eventually proves the presence of collagen triple helices in collagen suspension. In order to investigate the protein organisation following fibre formation, CD was employed. Type I collagen has a unique CD spectrum in which a small positive peak related to triple helix conformation appears at about 210 – 230 nm, a crossover near 213 nm and a large negative peak related to random coil conformation at around 197 nm [46, 52, 53]. Besides the in-house extracted rat tail collagen, these features were remarkably identified in the far-UV CD plots of wet-spun fibre acidic solutions. Fig 5 shows wet spun fibres having a positive peak at 220 nm and a large trough at 197 nm with a crossover at 214 nm; and these peaks correlated well to in-house extracted rat collagen. Moreover, the positive to negative peak ratio (RPN) in CD spectra of collagen triple helices and native fibres were 0.117 and 0.111, respectively. These RPN values suggest nearly complete retention (94%) of triple helices in the resulting wet-spun fibres. Whereas previous study has been reported that electrospinning of collagen with 1,1,1,3,3,3-hexafluoro-2-propanol (HFP) as a solvent produced only 55% retention of the triple helix structure [20]. Such observed nearly-preserved triple helicity in wet-spun fibres can be expected to primarily influence tensile and elastic properties, mineralisation capability and interaction with cells [1].

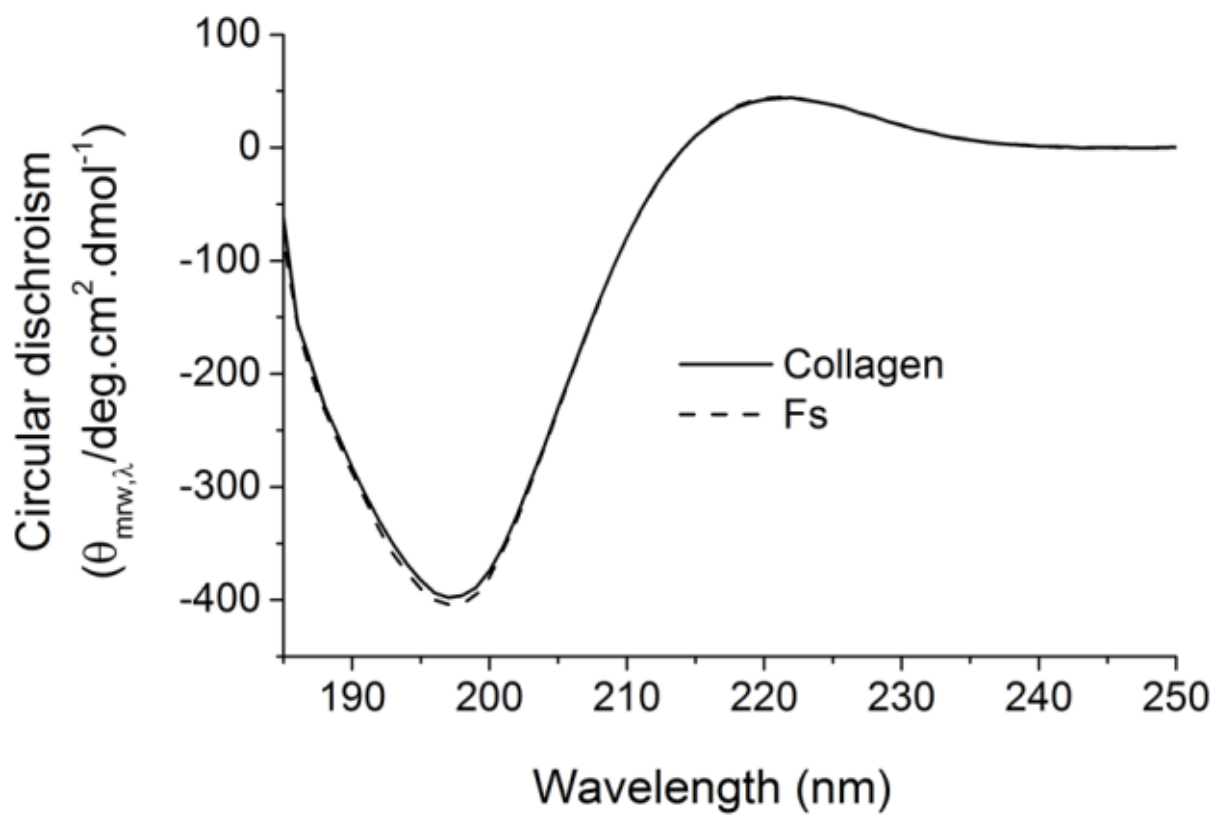

**Fig. 5.** Far-UV spectra of raw collagen triple helices and wet-spun fibres showing retention of the triple helix characteristic peak after fibre formation.

3.2 *Crosslinking fibres*

Wet spun fibres were stabilized using three different crosslinking methods with the aim of investigating whether the segment length of resulting crosslinking junctions, i.e. zero-length in the case of EN, low molecular weight junction in the case of Ph and high molecular weight junction in the case of PEG, could impact mechanical properties of resulting fibres. As shown in scheme 1, either Ph or PEG was NHS activated and subsequently reacted with free amino terminations (predominantly lysines) of collagen, leading to the formation of hydrolytically cleavable amide bonds between collagen molecules. The degree of collagen crosslinking in reacted fibres was quantified via TNBS colorimetric assay, enabling the determination of the molar content of free amino groups in collagen samples, being amino groups expected to primarily take part in the crosslinking reaction with NHS-activated diacid [48]. For each crosslinker, two different concentrations were studied to verify whether crosslinking formulation influenced the degree of collagen crosslinking as well as the mechanical, swelling and degradation properties of reacted fibres.

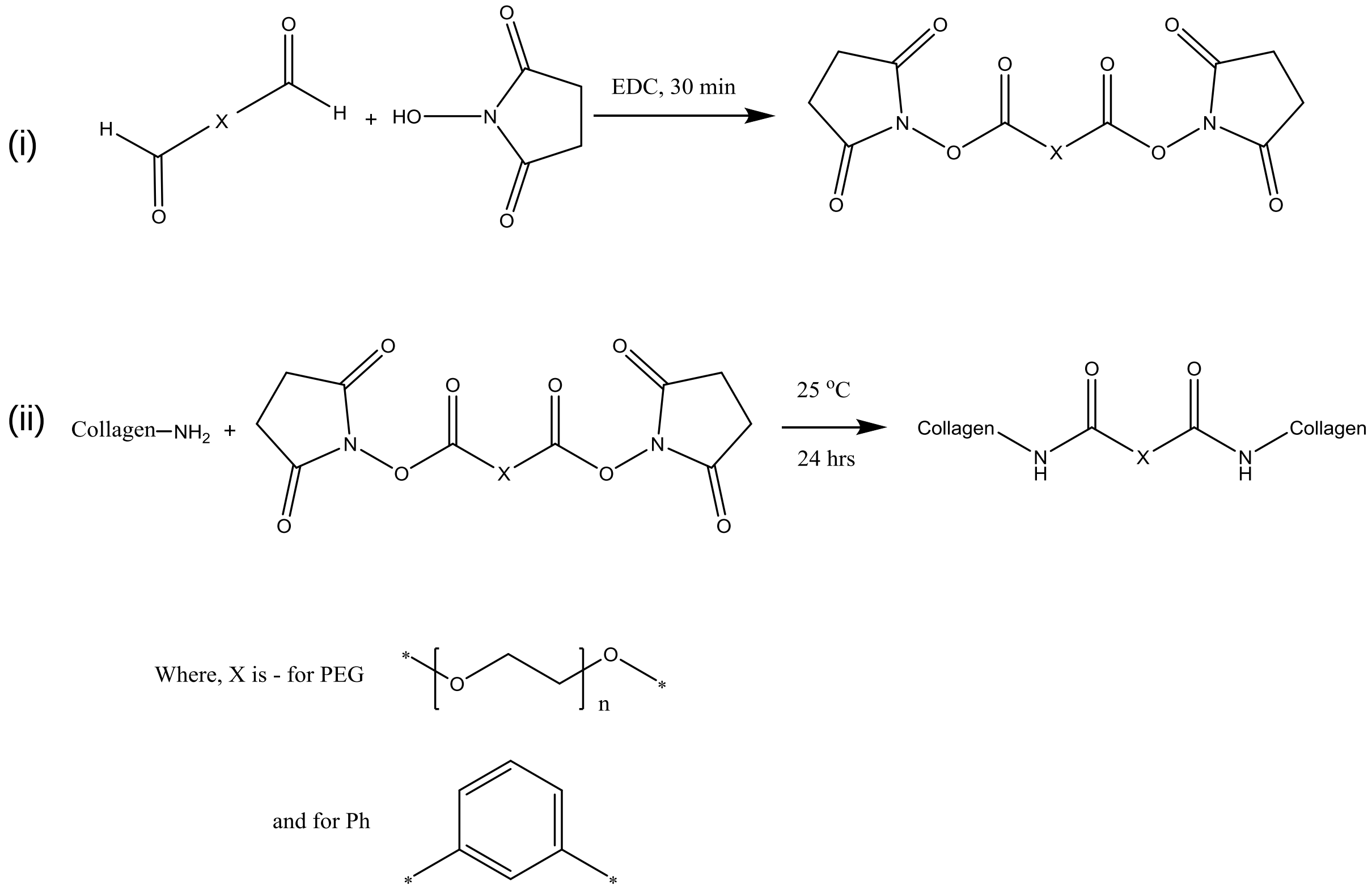

**Scheme 1.** Mechanism of the crosslinking reaction with varied diacid: (i) either PEG or Ph is NHS activated in presence of EDC, so that the crosslinking reaction of collagen can take place (ii). The functionalization of free amino terminations (mainly collagen lysines and amino termini) occurs through nucleophilic addition with activated carboxylic functions, leading to the formation of a covalent network of collagen triple helices.

Fig 6(a) shows the tensile modulus of fibres in both dry and hydrated states. All the crosslinked groups exhibited higher tensile modulus and strength compared to the non-crosslinked group. Among the crosslinked fibres, PEG-based materials displayed remarkably high modulus in both the dry and hydrated states, followed by Ph- and EDC-based materials. Statistically significant differences were obtained for collagen fibres crosslinked with high molecular weight junctions. The F-PEG0.5 and F-Ph0.5 groups exhibited tensile moduli of 1.5 and 1.2 times higher and tensile strengths of 1.3 times higher compared to the original fibres, respectively (Fig. S3). Thus, the segment length of the crosslinking junction between collagen molecules appeared to directly affect the tensile modulus of resulting fibres. This observation supports the hypothesis that crosslinking junctions with increased segment length are more likely to bridge distant collagen molecules in comparison with crosslinking junctions of decreased segment length, leading to materials with superior macroscopic properties. Fig 6(b) shows the typical stress-strain graph of Fs and F-PEG0.5 in both dry and hydrated state. F-PEG0.5 samples displayed up to 16% and 12% tensile strain in the dry and hydrated states, respectively, whereby in the hydrated state, no yield point could be observed in the control fibre samples. This suggests that the incorporation of crosslinking junctions among collagen molecules successfully stiffens the fibre.

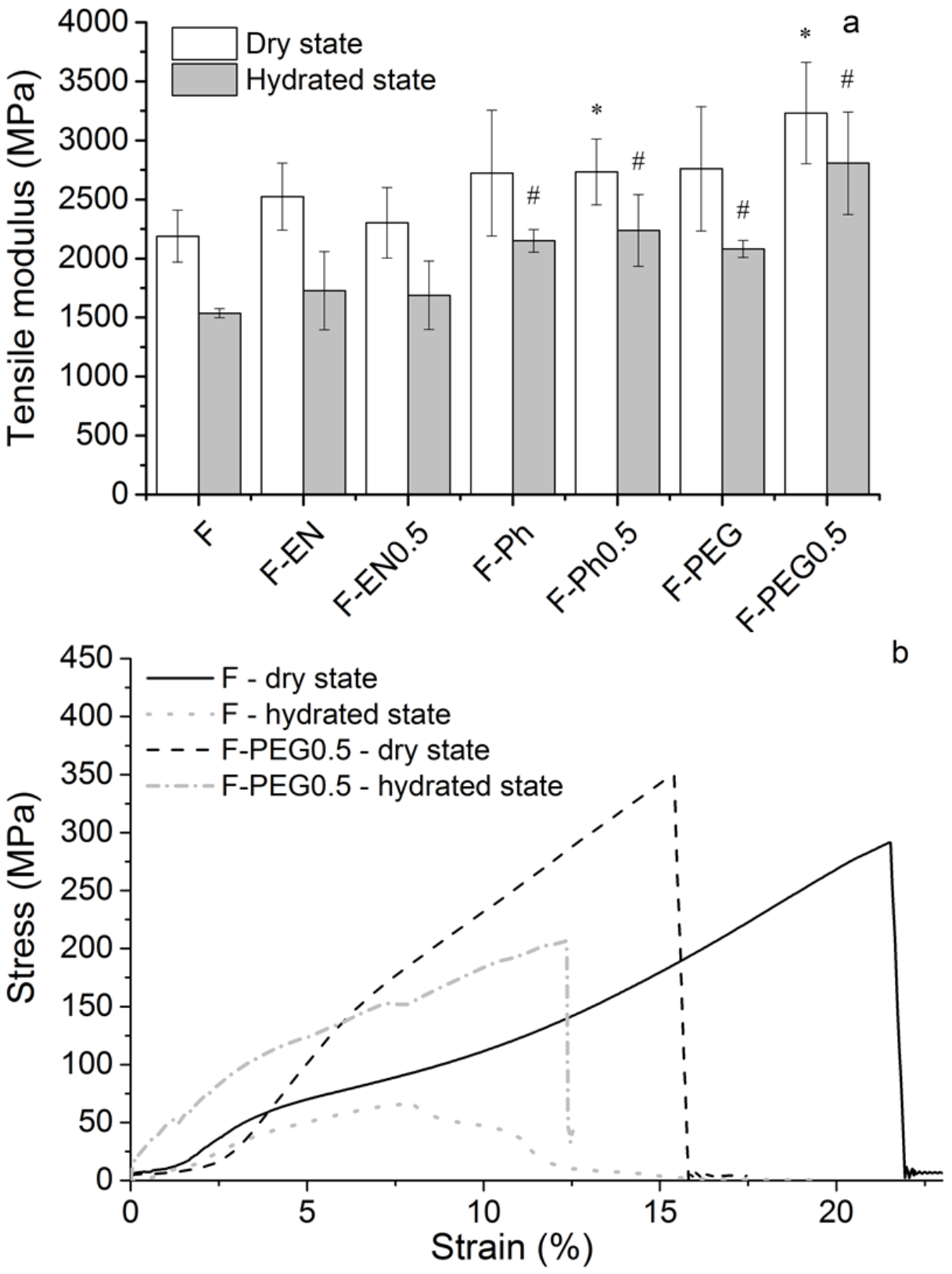

**Fig. 6.** Tensile modulus (a), and corresponding stress-strain curves (b) of either wet-spun or crosslinked fibres in both dry and hydrated states. ‘*’ and ‘#’ indicate that corresponding mean values in crosslinked samples are significantly different ($p < 0.05$, t-test) from the mean values of the wet-spun control group in the dry and hydrated states, respectively.

In the hydrated state, the F-PEG0.5 group showed 13% reduction in dry tensile modulus, in comparison to the 30% reduction observed in Fs. This variation in tensile properties is also supported by the fibre swelling study (table 1), where the F-PEG0.5 group showed only 69 wt.-% swelling after 1 day incubation in PBS at 37 ℃. The superior tensile properties of sample F-PEG0.5 may be attributed to the long segment length of the PEG crosslinker, likely promoting crosslinking of distant collagen molecules [33]. In comparison to F-PEG0.5, the group F-PEG displayed lower tensile modulus and strength. This observation likely indicates that increasing the concentration of PEG-based crosslinker during the crosslinking reaction does not directly affect the crosslink density of the collagen network, whilst a PEG-induced plasticisation effect of the collagen triple helices becomes predominant.

**Table 1**: Swelling ratio ($S_R$), mass change ratio ($M_R$), denaturation temperature ($T_d$) and degree of crosslinking ($C$) as determined on wet-spun and crosslinked fibres.

| Sample ID | $S_R$/wt% | $M_R$/wt% | $T_d$ (℃) | C/mol% |
|---|---|---|---|---|
| F-EN | 73 ± 7 | 18 ± 2 | 63 | 92 ± 1 |
| F-EN0.5 | 78 ± 1 | 13 ± 6 | 55 | 88 ± 2 |
| F-Ph | 91 ± 11 | 11 ± 3 | 60 | 93 ± 1 |
| F-Ph0.5 | 76 ± 11 | 12 ± 5 | 63 | 94 ± 1 |
| F-PEG | 96 ± 3 | 19 ± 7 | 60 | 91 ± 1 |
| F-PEG0.5 | 69 ± 8 | 22 ± 5 | 66 | 91 ± 1 |
| Fs | 293 ± 9 | - | 62 | - |

To further investigate the relationships between fibre tensile properties and the molecular architecture of the crosslinked triple helices, reacted samples were investigated via TNBS assay, hydrolytic degradation tests and DSC analysis. Table 1 shows that after 7 days of incubation in PBS at 37 ℃, the PEG modified group showed higher mass loss than the Ph- and EN-crosslinked samples, although there was no significant difference in mass loss with respect to the other fibre formulations. This reflects the fact that a degree of crosslinking of at least 88% was observed via TNBS assay and was only slightly affected among the different samples. Since the crosslinking junctions were introduced via amide net-points between collagen molecules, the comparable mass loss values likely reflected the fact that hydrolytic degradation mainly occurred via the cleavage of covalent bonds of comparable reactivity against water. The comparable degree of crosslinking between samples F-EN, F-Ph and F-PEG provide additional evidence that fibre mechanical properties were mostly ruled by the segment length of the introduced crosslinking junction. Considering the high degree of crosslinking and the low mass loss, minimal alteration in the morphology of F-PEG-0.5 was also observed via SEM image (Fig. S4) following 1-week incubation in PBS, indicating that resulting crosslinked fibres are wet stable in physiological conditions.

The thermal stability of fibres was studied using DSC, where the denaturation temperature ($T_d$) was measured as the endothermic peak associated with the unfolding of collagen triple helices into single poly-proline chains [46, 54-56]. Fig. S5 shows the thermograms of control fibres and cross-linked fibres. Endothermic thermal transitions were apparent in the range of 50 – 76 ℃ for all the groups. This is in agreement with previous studies where thermal denaturation of collagen was recorded at 55 – 67 ℃ [46, 57]. $T_d$ was recorded as 62 ℃ and 66 ℃ for Fs and F-PEG0.5, respectively, whereby the highest $T_d$ value was exhibited by sample F-PEG0.5, in line with previous tensile data. Endothermic peak at around 70 ℃ has also been noticed for other non-triple helical proteins; for example, *Bombyx mori* mulberry worm silk and soymilk protein showed an endothermic peak at 67 ℃ and 70 ℃, respectively, which in these cases was associated with either the rapid protein aggregation [58] or denaturation of 7S (β-conglycinin) [59], respectively. Because the F-PEG0.5 sample showed good thermo-mechanical properties in comparison to the other sample formulations, it was selected for further coating studies.

3.3 *Biomimetic CHA coating on F-PEG0.5*

The surface morphology of CHA coated fibres was characterized using SEM. As shown in Fig 7, F-PEG0.5-CHA exhibited rough fibre morphology, although a uniform coating was successfully applied across the whole surface. The coating had the appearance of globular apatite with a particle size of around 50 nm. The SEM images also showed no detrimental effect of the coating process on the morphology of the fibres, which is indicative of the effectiveness of the crosslinking process. These results could not be accomplished in the case of non-crosslinked fibres, whereby complete sample dissolution was observed following 24 hours incubation in PBS. This is obviously related to the lack of covalent crosslinks between collagen triple helices, resulting in unstable wet-spun fibres in aqueous environment.

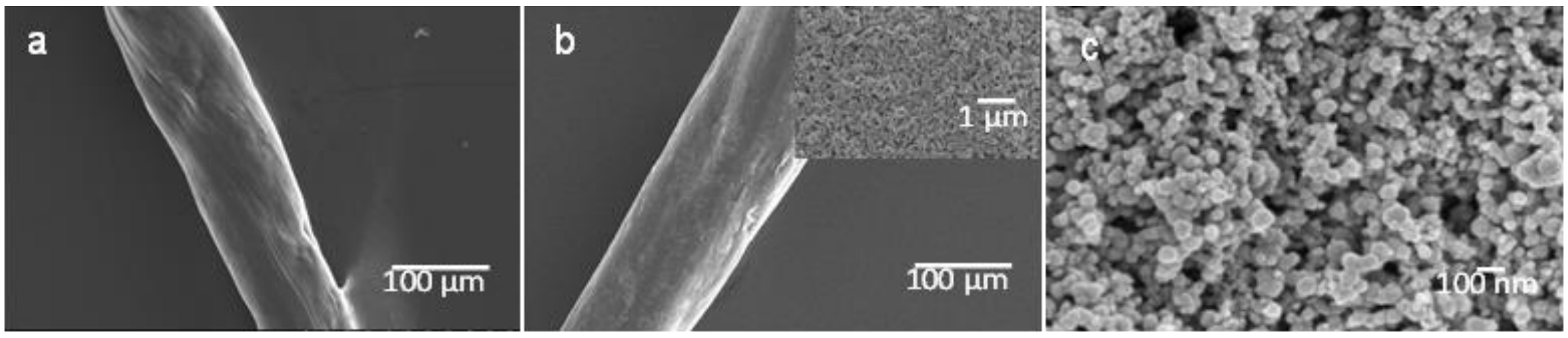

**Fig. 7.** SEM images of (a) F-PEG0.5; (b) F-PEG0.5-CHA; and (c) CHA coating at higher magnification.

Tensile testing data indicated that the dry tensile modulus of the F-PEG0.5-CHA fibres was 1.7 and 1.2 times higher than the Fs and F-PEG0.5, respectively, whilst the tensile strength was 1.7 and 1.3 times higher than the Fs and F-PEG0.5.

The chemical structure of the CHA coating was studied by ATR-FTIR. The FTIR spectrum of fibre in Fig 8 exhibits characteristic peaks at 1650, 1550 and 1240 $cm^{-1}$, attributable to amide I, II and III, respectively. The amide I absorption arises from the stretching vibrations of C=O groups, whereas the amide II is due to N-H bending and C-N stretching vibrations. The amide III is mainly the result of C-N stretching and N-H in-plane bending from amide linkages. Amide III can also be due to the wagging vibrations of $CH_2$ groups in the glycine backbone and proline side chains. Each of these amide bonds were detected without any shift in the FTIR spectra of both F-PEG0.5 and F-PEG0.5-CHA. The integrity of the collagen triple helix structure can be verified by the ratio of amide III to 1450 $cm^{-1}$ ($A_{III}/A_{1450}$), where a value of around 1 is expected for an intact triple helix structure and 0.5 for denatured collagen, i.e. gelatin [60]. The absorption ratio of $A_{III}/A_{1450}$ for Fs, F-PEG0.5 and F-PEG0.5-CHA was equal to or higher than 0.99. This result confirms that collagen triple helices were successfully preserved after crosslinking and the coating process, supporting the CD data observations. In the spectrum of F-PEG0.5-CHA two additional bands were observed at 1039 and 1400 $cm^{-1}$, which are characteristic bands for phosphate and carbonate groups, respectively. These bands are similar to those found in natural bone [61], and are similar to those previously reported for CHA coating [37].

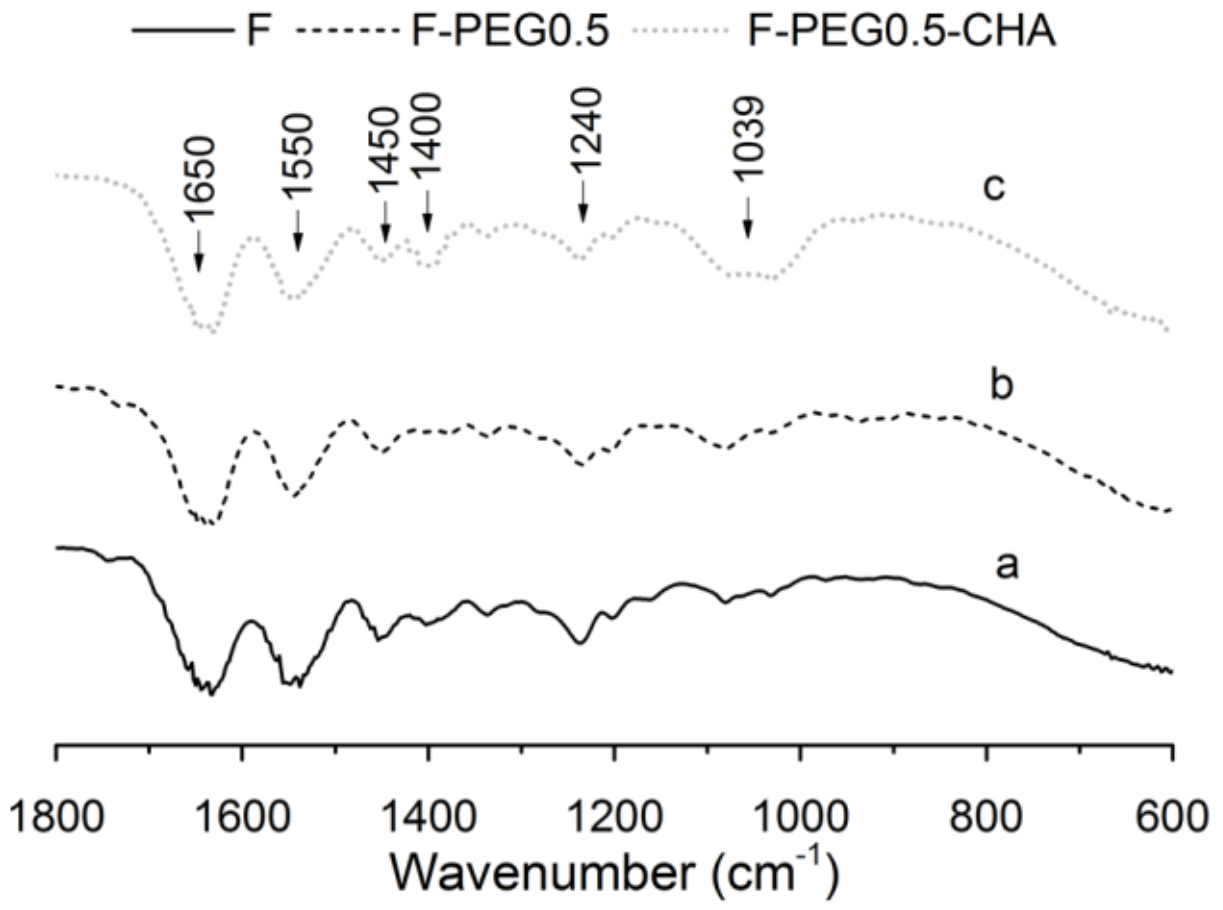

**Fig. 8.** ATR-FT-IR spectra of samples (a) F, (b) F-PEG0.5, and (c) F-PEG0.5-CHA. FTIR spectra (a) and (b) show characteristic peaks at 1650, 1550 and 1240 $cm^{-1}$, attributable to amide I, II and III, respectively. Additionally, FTIR spectra (c) exhibits two additional bands at 1039 and 1400 $cm^{-1}$, which are attributed to phosphate and carbonate groups, respectively.

The chemical structure of the coating was also studied by EDS, as shown in Fig 9. Characteristic calcium and phosphorus peaks were observed for the F-PEG0.5-CHA group, whereas there were no such characteristic peaks for Fs. Quantitative elemental analysis by EDS showed that the Ca/P atomic ratio for the coated group was 1.75, which is comparable to that previously found in CHA [37, 49].

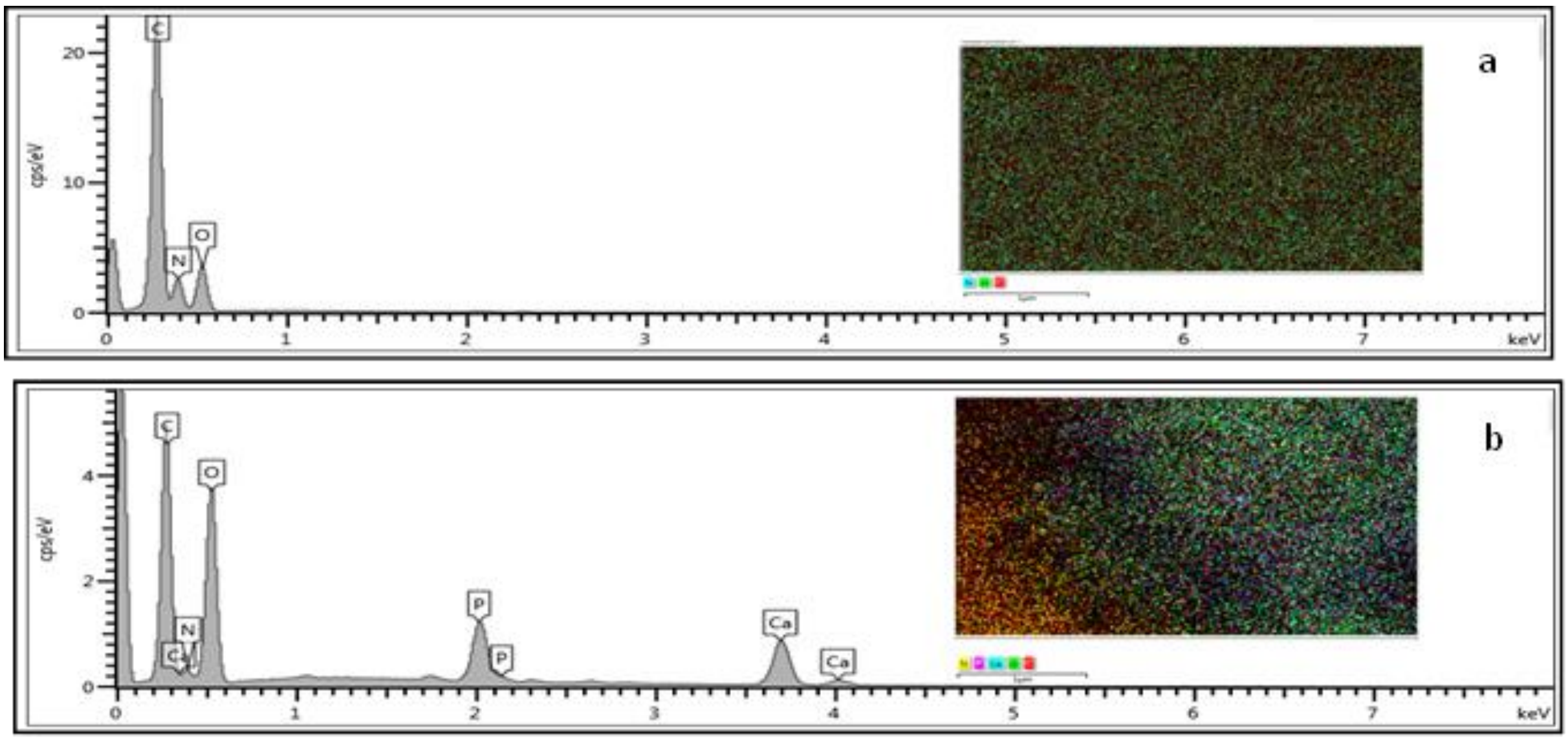

**Fig. 9.** EDS spectra of (a) Fs and (b) F-PEG0.5-CHA, where CHA coated F-PEG0.5 shows characteristic peaks for calcium and phosphorus.

3.4 *Cytotoxicity study*

The cyto-compatibility of fibres together with all the crosslinked fibres and coated F-PEG0.5-CHA samples was carried out by means of extract cytotoxicity assays to verify the potential tolerance of the fibres in a biological environment. As shown in Fig 10, mouse fibroblasts cell line L929 appeared confluent after 48 hr of culture on the extract for all the groups except the negative control group. The morphology observations are in line with MTS data, which shows all the groups are significantly different compared to the negative control group (Fig. 11). Other than that, there was no statistically significant difference among the crosslinked and coated groups, confirming that both crosslinking and coating reactions did not lead to the formation of toxic species, or to the presence of non-reacted, potentially toxic moieties, in the resulting materials.

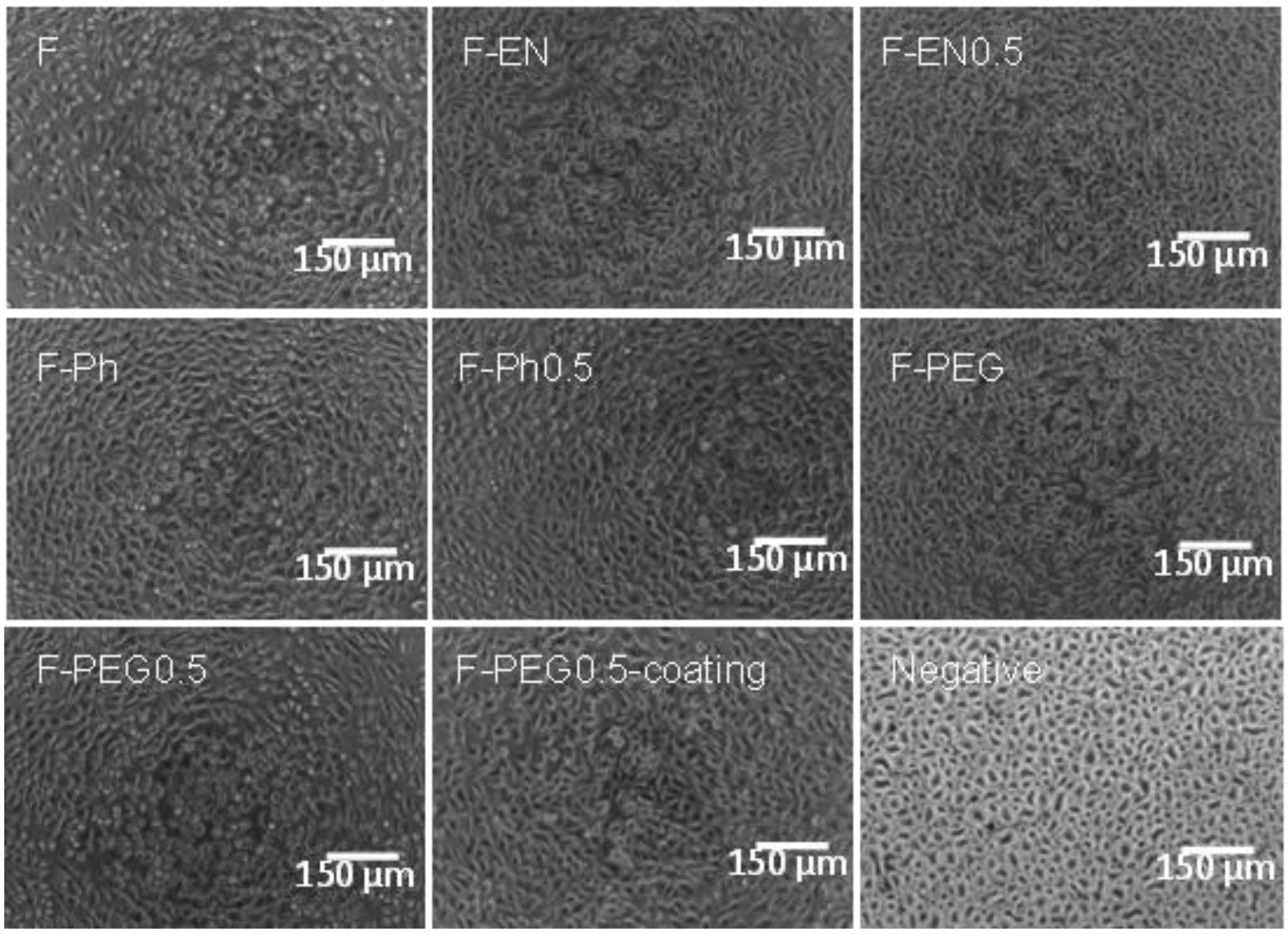

**Fig. 10.** Optical images of L929 mouse fibroblast cell morphology following 48 hr culture in sample extracts showing confluent cells for all the groups except the negative group.

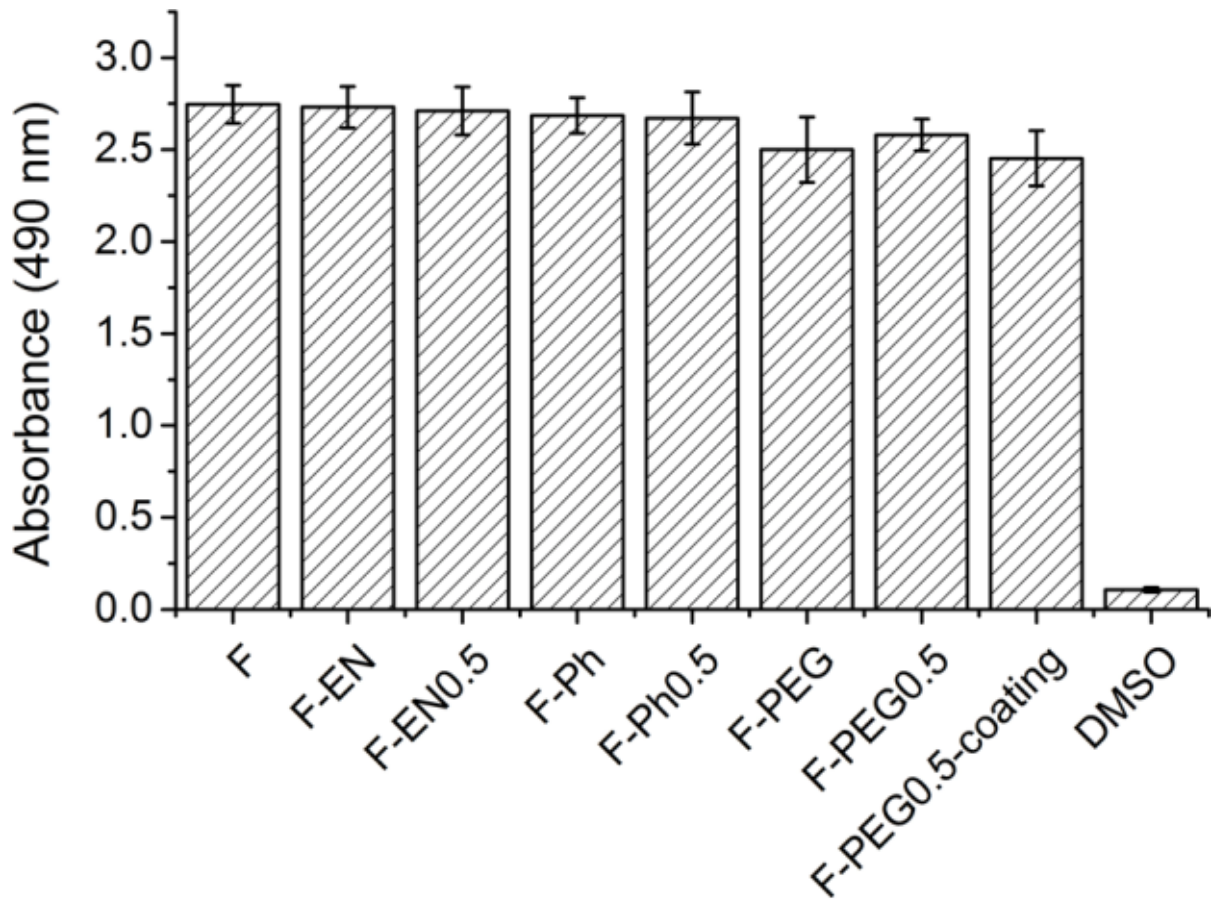

**Fig. 11.** Formazan absorbance following MTS assay on L929 cells after 48 hr cell culture showing all the studied groups (except DMSO) are comparable with no sign of cytotoxicity.

## 4. Conclusions

In this study, we have shown that biomimetic wet-stable fibres can be successfully manufactured via wet-spinning and diacid-based crosslinking of collagen triple helices. A concentration range of 1.2-1.6% (w/v) collagen was identified for the formation of wet-spun fibres with retained triple helix stability (~94%), homogeneous fibre morphology and remarkable tensile modulus (*E*: 400-2200 MPa). Wet-spun fibres were reacted with diacid-based crosslinkers of varied molecular weight, including both zero length and long range crosslinkers. The long range bifunctional crosslinker, PEG, showed markedly enhanced mechanical properties in both dry and hydrated states compared to both EN- and Ph-crosslinked fibres, suggesting the functionalisation of distant collagen triple helices at the molecular level. CHA was successfully coated on sample F-PEG0.5 through a biomimetic precipitation process, whilst no morphological changes were observed in the underlying fibre in light of the presence of the covalent collagen network. L929 cell culture on extracts for 48 hr revealed no sign of cytotoxicity in optical images or MTS assays. This study extends the potential use of wet spinning technology towards the design of biomimetic materials with customised architecture. The presented CHA-coated crosslinked fibres could be assembled into nonwoven fabrics and represent promising material systems for GBR, e.g. aiming at the repair of maxillofacial bone defects.

**Acknowledgement:**

This work is funded by the EPSRC Centre for Innovative Manufacture in Medical Devices (MeDe Innovation). The support of The Clothworkers' Centre for Textile Materials Innovation for Healthcare is also gratefully acknowledged. The authors would like to thank J. Hudson, M. Fuller and C. Gough for their help with SEM, mechanical testing and cell culture, respectively.

## Supplementary document:

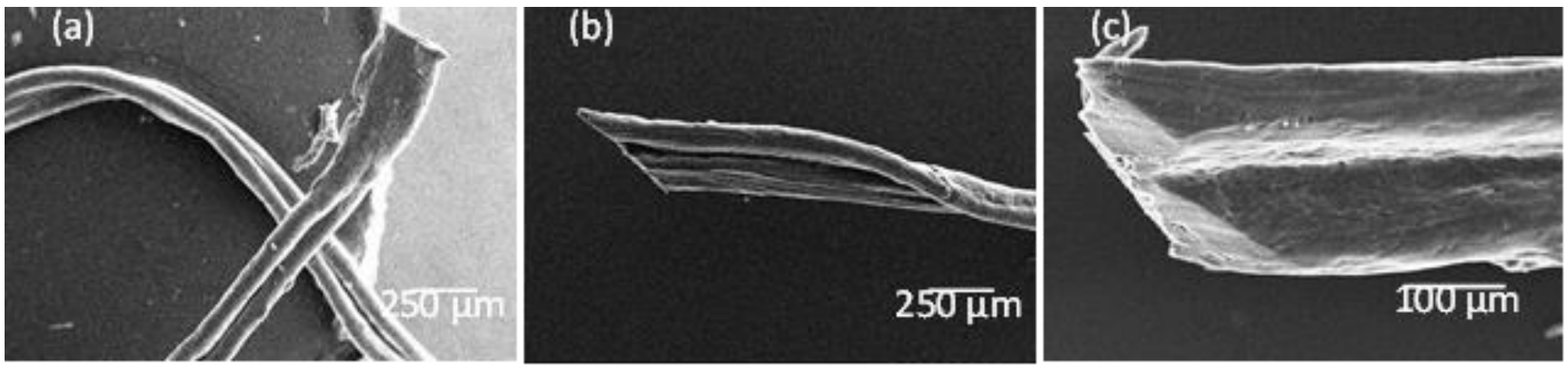

**Fig. S1.** SEM images of fibres wet-spun from collagen suspensions with 1.6% collagen (w/v) concentration showing surface striation (a) along the fibre and (b,c) at the fibre edges.

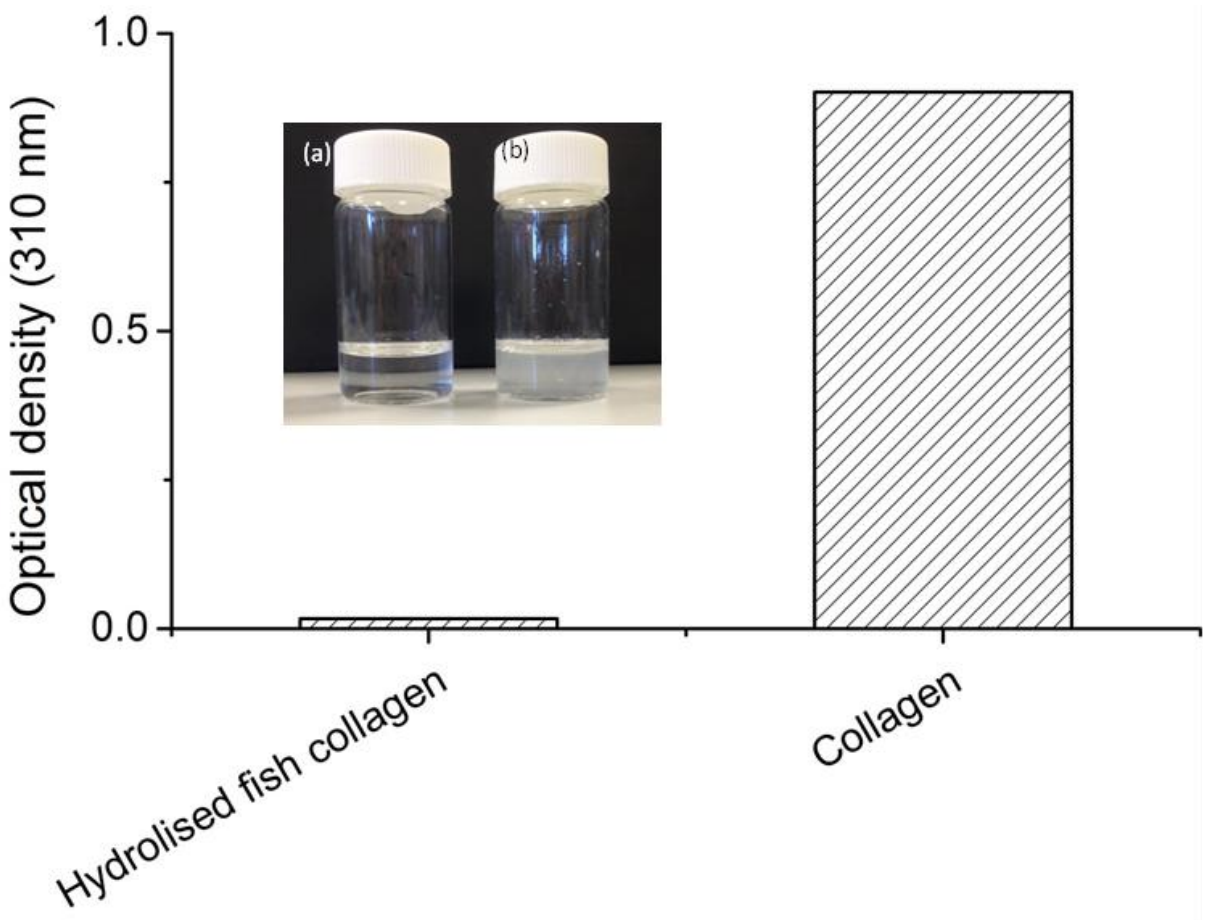

**Fig. S2.** Optical density of dissolved hydrolised fish collagen and in-house extracted collagen. In inset optical image of 0.25% (w/v) of (a) hydrolyzed fish collagen and (b) collagen in 17.4 mM acetic acid solution. Both the graph and the optical image are in correlation and showed much higher optical density of collagen, thereby indicating that the collagen is present as a triple helix in the collagen suspension.

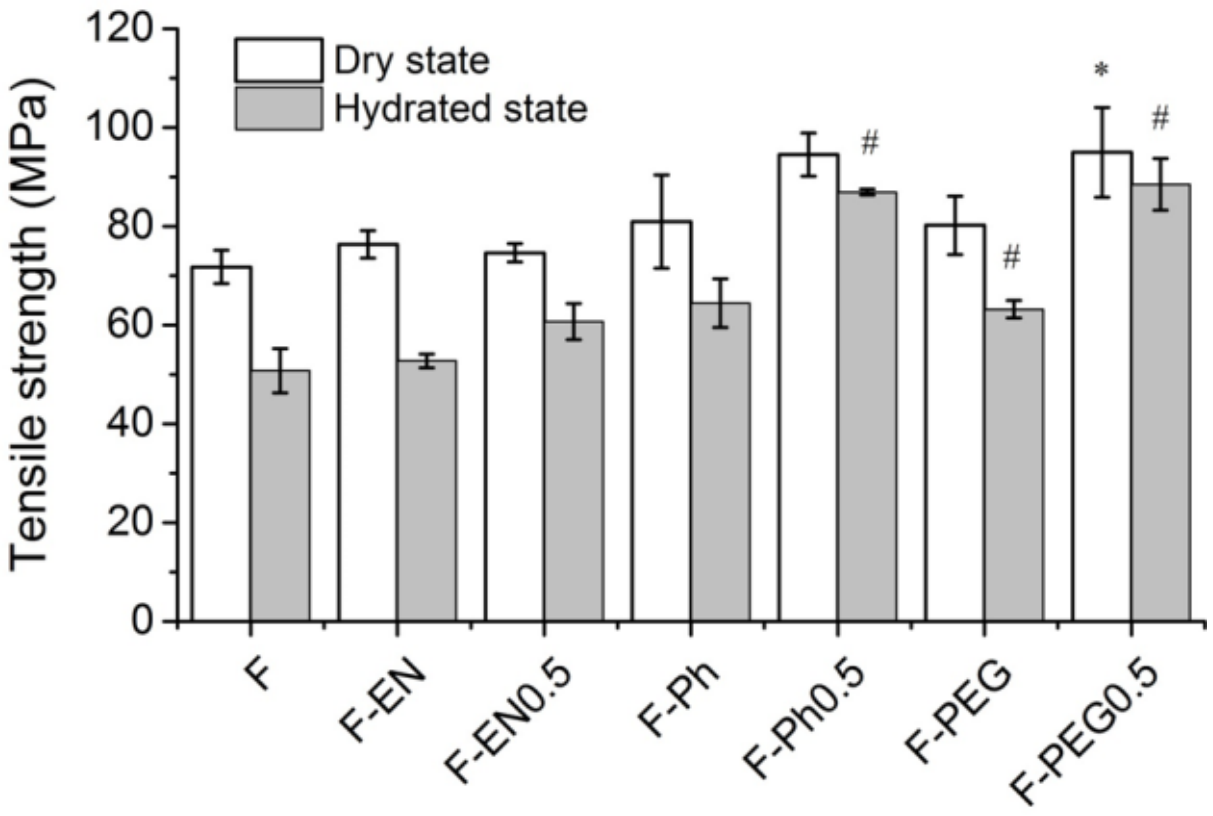

**Fig. S3.** Tensile strength of wet-spun and crosslinked fibres. ‘*’ and ‘#’ indicate that mean values of corresponding crosslinked samples are significantly different from the wet-spun control group in dry and hydrated state, respectively ($p < 0.05$, t-test).